# A SLIDING MODE CONTROL FOR A SENSORLESS TRACKER: APPLICATION ON A PHOTOVOLTAIC SYSTEM

Ahmed Rhif [1,2]

1. Advanced System Laboratory, Polytechnic School of Tunisia
2. Department of Electronic, High Institute of Applied Sciences and Technologies, Sousse, Tunisia

E-mail: ahmed.rhif@gmail.com

## ABSTRACT

*The photovoltaic sun tracker allows us to increase the energy production. The sun tracker considered in this study has two degrees of freedom (2-DOF) and especially specified by the lack of sensors. In this way, the tracker will have as a set point the sun position at every second during the day for a period of five years. After sunset, the tracker goes back to the initial position (which of sunrise). The sliding mode control (SMC) will be applied to ensure at best the tracking mechanism and, in another hand, the sliding mode observer will replace the velocity sensor which suffers from a lot of measurement disturbances. Experimental measurements show that this autonomic dual axis Sun Tracker increases the power production by over 40%.*

## 1. INTRODUCTION

The first automatic solar tracking system was presented by McFee [1-2]. For this, author had developed an algorithm to compute the flux density distribution and the total received power in a central receiver solar power [2]. Few years later, Semma and Imamru used a microprocessor to adjust the solar collectors' positions in a photovoltaic concentrator such that they are pointed toward the sun at every moment [1]. Since 1881, knowledge has evolved and we know now how to capture the solar energy. In this way, the use of the artificial solar energy is currently processed in three different processes: the photothermal conversion that converts the radiation to usable heat (heat engines, water heaters ...), the photovoltaic conversion that converts radiation into electrical current and the photochemical conversion that allows the store of thermal energy by the deformation of molecules. For each of these sectors, energy is received by sensors which are perpendicular to the solar radiation which will increase the amount of the collected energy collected during the day.  It has been shown experimentally we kept the sensor perpendicular to the solar radiation; the flat plate photovoltaic panel had an average gain of 20% compared with a horizontal sensor.

More sophisticated sensors than flat plate collectors such as the concentration sensors have been emerged as well as in the thermal photovoltaic [3].

A sun tracker is a process which orients various payloads toward the sun such as photovoltaic panels, reflectors, lenses… or other optical devices.

The sun trackers of photovoltaic panels are used to minimize the incidence angle between the photovoltaic cell and the incoming light the fact that increases the amount of the produced energy by about 40%. In photovoltaic plants, it is estimated that solar trackers take places in at least 85% of new commercial installations, in 2009 to 2011, when production is greater than 1MW.
There are different types of trackers:

✓ Passive tracker: the passive trackers use a boiling point from a compressed fluid that is driven to one side to other by the solar heat which creates a gas pressure that may cause the tracker movement [1]. As this process presents a bad quality of orientation precision, it turns out to be unsuitable for certain types of photovoltaic collectors. The term passive tracker is used too for photovoltaic panels that include a hologram behind stripes of photovoltaic cells. In this way, sunlight reflects [2] on the hologram which allows the cell heat from behind, thereby increasing the modules' efficiency. Moreover, the plant does not have to move while the hologram still reflects sunlight from the needed angle towards the photovoltaic cells.

✓ Chronological tracker: the chronological tracker counteracts the rotation of the earth by turning in the opposite direction at an equal rate as the earth [3]. Actually the rates are not exactly equal because the position of the sun changes in relation to the earth by 360° every year. A chronological tracker is a very simple, potentially and very accurate solar tracker specifically used with polar mount. The process control is ensured by a motor that rotates at a very slow rate (about 15 degrees/ hour).

✓ Active tracker: the active trackers use two motors (Fig. 1) and a gear trains to drive the tracker by a controller matched to the solar direction [4]. In fact, two axis trackers are used to orient movable mirrors of heliostats that reflect the sunlight toward the absorber of a power station central [5]. As each mirror will have an individual orientation, those plants are controlled through a central computer system which allows also the system, when necessary, to be shut down.

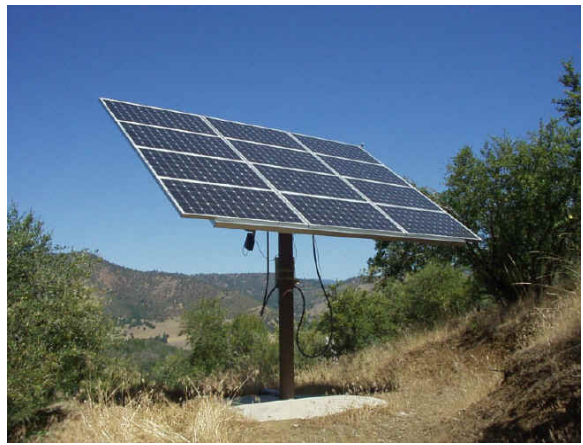

Figure 1. Two Axis Solar Tracker

## 2. THE DUAL AXIS SUN TRACKER (DAST)

In areas where sunlight is weak, or in those areas where subsidies are not very expanded, the initial cost of solar panels can be extremely high. A solution to improve the energy production exists: the solar tracker, or the sun tracker. The sun tracker allows placing the panel in relation to the best position of the sun (orthogonal to radiation if possible). Indeed, the position of the sun varies constantly, both during the day and during different times of the year. The perfect solution is to use a solar tracker of six axes to track the exactly sun movements. This type of system requires a minimum of clearance to allow free movement in all directions. It is therefore incompatible with an integrated roof, except perhaps for systems with two axes. In 1992, Agarwal presented a two axis tracking process consisted of a worm gear drives and four bar type kinematic linkages that make easy the focusing operation of the reflectors in a solar concentrator system [6]. A Dual Axis Sun Tracker (Fig.2) has two degrees of freedom that act as axes of rotation. It has a vertical axis (primary axis) perpendicular to the ground and a horizontal one typically normal to the primary axis. The first axis is a vertical pivot shaft that allows the device to move to a compass point. The second one is a horizontal elevation pivot implanted upon the platform. To adjust for sun azimuth, azimuth motor drives the vertical axis that makes the panel rotates [1, 5, 7]. To adjust for the sun altitude, a second motor will operate for the panel elevation acting on the horizontal axis.

Using the combinations between the two axes (azimuth and altitude), any location in the upward hemisphere could be pointed. Such systems should be operated under a computer control or a microcontroller according to the expected solar orientation. Also, it may use a tracking sensor to control motors that orient the panels toward the sun. This type of plant is also used to orient the parabolic reflectors that mount a sterling engine to produce electricity at the device. In order to control the movement of these plants, special drives are designed and rigorously tested. For that, many considerations during cloudy periods must be expected to keep the tracker out from wasting energy.

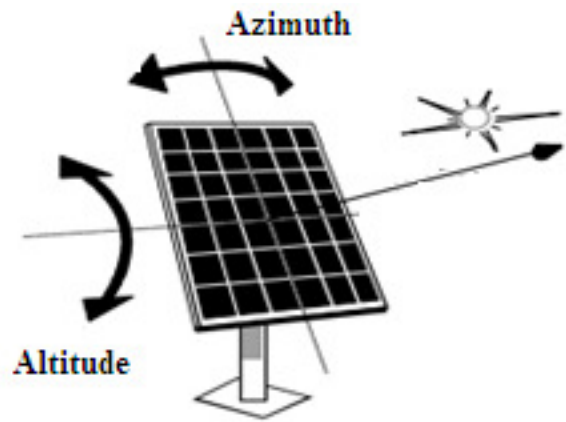

Figure 2. The DAST system orientation

In fact after tests, a solar panel of 3 m² fixed to a surface produces about 5 kWh of electricity per day. The same installation, but equipped with a tracker, can provide up to 8 kWh per day. In conclusion, this device allows the increase of the produced energy amount by about 40% in relation to the fixed panels (Fig. 3). So, the tracker presents a very good tool for energy production optimization.

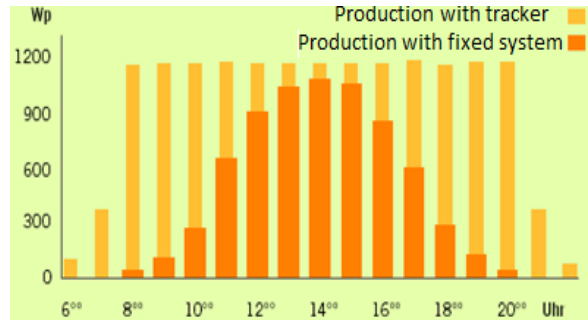

Figure 3. Histogram of the amount energy produced

If we consider that the system operation is accomplished by one of those stepper motors at each time, the mathematical model of the process in the (d-q) referential "**d**irect-**q**uadrature transformation"[6] (Fig.4) would be represented in (1).

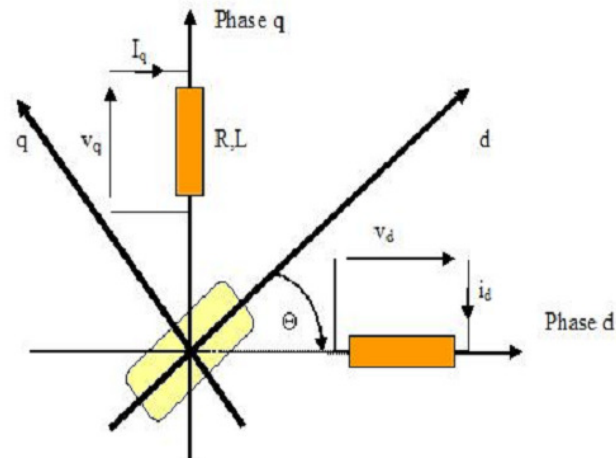

Figure 4. The electrical model in the (d-q) referential

$$
\begin{cases}
\dfrac{di_d}{dt} = \dfrac{1}{L}\left(v_d - Ri_d + NL\Omega\ i_q\right) \\[2mm]
\dfrac{di_q}{dt} = \dfrac{1}{L}\left(v_q - Ri_q + NL\Omega\ i_d - K\Omega\ \right) \\[2mm]
\dfrac{d\Omega}{dt} = \dfrac{1}{J}\left(Ki_q - f_v\Omega\ - C\ \right) \\[2mm]
\dfrac{d\theta}{dt} = \Omega
\end{cases}
\qquad (1)
$$

with $R$ resistance, $L$ inductance, $J$ inertia, $N$ number of spin, $K$ torque constant gain, $f_v$ friction, $\Omega$ angular velocity, $\theta$ the rotor position, $i_d$ and $i_q$ current in the (d-q) referential, $v_d$ and $v_q$ voltage in the (d-q) referential.

Consider,

$$\begin{cases} x = \left[i_d, i_q, \theta, \Omega\right]^T \\ u = \left[v_d, v_q\right]^T \end{cases}$$

then the model state function will be represented in (2).

$$\dot{x} = f(\Omega)x + gu + p \qquad (2)$$

Where

$$f(\Omega) = \begin{bmatrix} -\dfrac{R}{L} & N\Omega & 0 & 0 \\ N\Omega & -\dfrac{R}{L} & a_{23} & -K \\ 0 & 0 & 0 & 1 \\ 0 & \dfrac{K}{J} & 0 & -\dfrac{f_v}{J} \end{bmatrix} \quad g = \begin{bmatrix} \dfrac{1}{L} & 0 \\ 0 & \dfrac{1}{L} \\ 0 & 0 \\ 0 & 0 \end{bmatrix} \quad p = \begin{bmatrix} 0 \\ 0 \\ 0 \\ -\dfrac{C}{J} \end{bmatrix}$$

With $f$ and $g$ are linear functions.

Other ways, this system (2) could be represented in a simplest way which makes much easy the control study. For that, we consider $y_1 = \theta$ and $y_2 = i_d$ so system (1) gives:

$$\begin{cases} \theta = y_1 \\ \Omega = \dot{y}_1 \\ i_d = y_2 \\ i_q = \dfrac{1}{K}\left(J\ddot{y}_1 + f_v \dot{y}_1 + C\right) \\ v_d = L\dot{y}_2 + Ry_2 - \dfrac{NL}{K}\dot{y}_1\left(J\ddot{y}_1 + f_v \dot{y}_1 + C\right) \\ v_q = \dfrac{JL}{K}y_1^{(3)} + \dfrac{1}{K}\left(Lf_v + RJ\right)\ddot{y}_1 + \left(\dfrac{Rf_v}{K} + K + NLy_2\right)\dot{y}_1 + \dfrac{L}{K}\dot{C} \end{cases} \qquad (3)$$

## 3. CONTROLLER DESIGN

Literature presents many types of controller which was tested and implemented in the sun tracker. In [1], Hua and Shen compared the sun tracking efficiencies of Maximum Power Point Tracking (MPPT) algorithms and used a control method which combined a discrete control scheme and a Proportional Integral (P.I) controller to track the maximum power points of a solar array [8-10]. In [12], Yousef developed a sun tracker system in which a fuzzy logic control implemented on a PC controlled the nonlinear dynamics of the tracking mechanism. For each tracking system, the modelling and controller design were done using a first order fuzzy inference system [3]. Moreover, Roth [11] have designed and implemented a solar tracking system using a pyrheliometer to measure the direct solar radiation. The system was controlled by a closed loop servo system composed by four quadrant photo-detectors needed to sense the position of the sun [12-13]. Finally in this section we will talk about a new control test using the sliding mode control which is very solicited in tracking problems.

## 3.1 The Sliding Mode Control

The appearance of the sliding mode approach occurred in the Soviet Union in the Sixties with the discovery of the discontinuous control and its effect on the system dynamics. This approach is classified in the monitoring with Variable System Structure (VSS) [14-16]. The sliding mode is strongly requested knowing its facility of establishment, its robustness against the disturbances and models uncertainties [17-18]. The principle of the sliding mode control is to force the system to converge towards a selected surface and then to evolve there in spite of uncertainties and the disturbances. The surface is defined by a set of relations between the state variables of the system. The synthesis of a control law by sliding mode includes two phases:

▪ the sliding surface is defined according to the control objectives and to the wished performances in closed loop,

▪ the synthesis of the discontinuous control is carried out in order to force the trajectories of the system state to reach the sliding surface, and then, to evolve in spite of uncertainties, of parametric variations,… the sliding mode exists when commutations take place in a continuous way between two extreme values $u_{max}$ and $u_{min}$. For any control device which has imperfections such as delay, hystereses, which impose a frequency of finished commutation, the state trajectory oscillate then in a vicinity of the sliding surface, a phenomenon called chattering appears [19].

## 3.2 The High Order Sliding Mode Control

The high order sliding mode consists in the sliding variable system derivation [20-21]. This method allows the total rejection of the chattering phenomenon while maintaining the robustness of the approach. For this, in the twisting algorithm; the system control is increased by a nominal control $u_e$. If we derive the sliding surface (S) n times we see that the convergence of S is even more accurate when n is higher. Other ways, the super twisting algorithm: the system control is composed of two parts $u_1$ and $u_2$ with $u_1$ equivalent control and $u_2$ the discontinuous control used to reject disturbances. In this case, there is no need to derive the sliding surface. To obtain a sliding mode of order n, in this method, we have to derive the error of the system n times.

In the literature, different approaches have been proposed for the synthesis of nonlinear surfaces [20-24]. In [22], the proposed area consists of two terms, a linear term that is defined by the Herwitz stability criteria and another nonlinear term used to improve transient performance.
In [23], to measure the armature current of a DC motor, Zhang Li used the high order sliding mode since it is faster than traditional methods such as vector control ... To eliminate the static error that appears when measuring parameters we use a P.I controller [24]. Thus the author have chosen to write the sliding surface in a transfer function of a proportional integral form while respecting the convergence properties of the system to this surface. The same problem of the static error was treated by adding an integrator block just after the sliding mode control.

The tracking problem of the photovoltaic panel is treated by using sliding mode control with nonlinear sliding surface as shown in (4).

$$s(t) = k_1\,e(t) + k_2\,\dot{e}(t) \qquad (4)$$

with $e(t)$ is the system error, $k_1 > 0$ and $k_2 > 0$.
To ensure that the state convergence to the sliding hyperplane, we have to verify the Lyaponov stability criterion (5) [25-26].

$$s\dot{s} \leq -\eta |s| \qquad (5)$$

$$s(x) = \dot{s}(x) = 0 \qquad (6)$$

with $\eta > 0$.

The control law can be then composed of two parts:

$$u = u_0 + u_1 \qquad (7)$$

- ▪ $u_0$ the nominal control
- ▪ $u_1$ the discontinuous control allowing to reject the disturbances.

To deal with the path tracking, we have, first, to stabilize the tracking error in the origin. For that, we consider:

$$e = \left[i_d - i_{dr}, i_q - i_{qr}, \Omega - \Omega_r, \theta - \theta_r\right]^T = \left[e_1, e_2, e_3, e_4\right]^T$$

Using system (1) we get (8).

$$\begin{cases} \dot{e}_1 = \dfrac{1}{L}\left(v_d - v_{dr} - \mathrm{Re}_1 + NL(e_3 e_2 + e_3 i_{qr} + e_2 \Omega_r)\right) \\[2mm] \dot{e}_2 = \dfrac{1}{L}\left(v_q - v_{qr} - \mathrm{Re}_2 + NL(e_3 e_1 + e_3 i_{dr} + e_1 \Omega_r) - Ke_3\right) \\[2mm] \dot{e}_3 = \dfrac{1}{J}\left(Ke_2 - f_v e_3 - C_r\right) \\[2mm] \dot{e}_4 = e_3 \end{cases} \qquad (8)$$

with $e$ the system error, $i_{dr}$ and $i_{qr}$ the reference current in the (d-q) referential, $v_{dr}$ and $v_{qr}$ the reference voltage in the (d-q) referential, $\Omega_r$ reference angular velocity, $\theta_r$ the reference rotor position.

Now, we start to study the velocity control by a second order sliding mode control. The sliding surface used to ensure the existence of this approach is written in (9).

$$s_\Omega = \mu e_3 + \dot{e}_3 \qquad (9)$$

with $\mu > 0$.

The first order derivative of the equation (9) gives:

$$\dot{s}_\Omega = \mu \dot{e}_3 + \ddot{e}_3$$

$$\dot{s}_\Omega = \mu \frac{1}{J}\left(Ke_2 - f_v e_3 - C_r\right) + \frac{1}{J}\left(K\dot{e}_2 - f_v \dot{e}_3 - \dot{C}_r\right)$$

In the second phase, we have now to deal with the position control process. In this way, we consider the sliding surface (10).

$$s_\theta = \mu_1 e_4 + \mu_1 \dot{e}_4 + \ddot{e}_4 \qquad (10)$$

Eqs (7) and (10) give:

$$s_\theta = \mu_1 e_4 + \mu_2 e_3 + \frac{1}{J}\left(Ke_2 - f_v e_3 - C_r\right)$$

with $\mu_1, \mu_2 > 0$

In the convergence phase, to bring the system on the sliding surface, different forms of switching control are available. Here we choose the following one represented in (11) and (12).

$$u_{s1} = -U_0 sign(s_\Omega) \tag{11}$$

$$u_{s2} = -U_0 sign(s_\theta) = \dot{s}_\theta \tag{12}$$

with $U_0 > 0$.

## 4. THE SLIDING MODE OBSERVER

The angular velocities of the stepper motors measurements are a hard task and could give unreliable results because of the disturbances that influence the tachymetry sensor. In this way, it may be adequate to estimate the angular velocity $\Omega$ using an observer instead of the sensor. This operation would give better results and will reduce the number of the sensors which are very costly and may have hard maintenance skills.

The observer can reconstruct the state of a system from the measurement of inputs and outputs. It is used when all or part of the state vector cannot be measured. It allows the estimation of unknown parameters or variables of a system. This observer can be used to reconstruct the speed of an electric motor, for example, from the electromagnetic torque. It also allows reconstructing the flow of the machine etc…

Sliding mode control can be used in the state observers design. These nonlinear high gain observers have the ability to bring coordinates of the estimator error dynamics to zero in finite time. Additionally, sliding mode observers (Fig.5) have an attractive measurement noise resilience which is similar to a Kalman filter. The chattering concerned with the sliding mode method can be eliminated by modifying the observer gain, without sacrificing the control robustness and precision qualities [27-30].

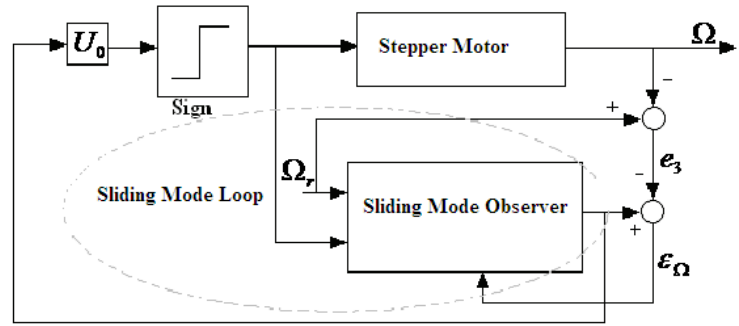

Figure 5. Velocity control based on a sliding mode observer

The observed position and angular velocity errors are noted respectively in (13) and (14).

$$\varepsilon_\theta = \theta - \hat{\theta} \tag{13}$$

$$\varepsilon_\Omega = \Omega - \hat{\Omega} \tag{14}$$

with $\hat{\theta}$ the estimated position and $\hat{\Omega}$ the estimated angular velocity.

Notice that the angular velocity is equal to the position derivative so we can write:

$$\dot{\varepsilon}_\theta = \frac{d\theta}{dt} - \frac{d\hat{\theta}}{dt} = \Omega - \hat{\Omega} = \varepsilon_\Omega$$

In this case, the system (1) could be represented as follow:

$$\begin{cases} \dfrac{d\hat{\Omega}}{dt} = \dfrac{K}{J}i_q - \dfrac{f_\nu}{J}\hat{\Omega} - \dfrac{C_r}{J} + \lambda_2 sign(\Omega - \hat{\Omega}) \\ \dfrac{d\hat{\theta}}{dt} = \hat{\Omega} + \lambda_1 sign(\theta - \hat{\theta}) \end{cases} \quad (15)$$

with $\lambda_1, \lambda_2 > 0$ and

$$\begin{cases} \dot{\varepsilon}_\theta = \varepsilon_\Omega - \lambda_1 sign(\varepsilon_\theta) \\ \dot{\varepsilon}_\Omega = \dfrac{d\Omega}{dt} - \dfrac{d\hat{\Omega}}{dt} = -\dfrac{f_\nu}{J}\varepsilon_\Omega \end{cases} \quad (16)$$

To make certain that the state converge to the sliding surface, we have to verify the Lyaponov stability criterion gave by the quadratic function $V_1 = \frac{1}{2}\varepsilon_\theta^2$.

In this way, $\dot{V}_1 = \varepsilon_\theta \left( \varepsilon_\Omega - \lambda_1 sign \left( \varepsilon_\theta \right) \right)$

then, $\dot{V}_1 < 0$ if $\lambda_1 > |\varepsilon_\Omega|_{\max}$ and the system dynamic is given by $\varepsilon_\Omega = \lambda_1 sign(\varepsilon_\theta)$ when $\varepsilon_\theta \to 0$

Consider now a second Lyaponov quadratic function $V_2 = \frac{1}{2}\left( \varepsilon_\theta^2 + J\varepsilon_\Omega^2 \right)$.

as $\varepsilon_\theta = 0$, system (14) gives: $\dot{V}_2 = -f_\nu \varepsilon_\Omega^2 - \varepsilon_\Omega(\lambda_2 sign(\varepsilon_\Omega) - C_r)$ then $\dot{V}_2 < 0$ if $\lambda_2 > |C_r|_{\max}$.

# 5. EXPERIMENTAL RESULTS AND DISCUSSIONS

Experimental results are accomplished for a second order sliding mode control of the state function (2) applied on the system shown in Figure 3 connected to a computer through a serial cable RS232. In this experience, we need to study the system control evolution, the real and observed angular velocities $\Omega_1$ and $\Omega_2$, the azimuth and the altitude inclinations $\theta_1$ and $\theta_2$. The sliding surface parameters used for this experience are $\mu = 0.135$, $\mu_1 = 1.2$ and $\mu_2 = 0.355$.

Table 1 shows the stepper motors characteristics.

## Table 1. Experimental parameters of the stepper motor

| | | |
|---|---|---|
| J | The inertia moment | $3.0145 \ 10^{-4} \text{Kg.m}^2$ |
| C | Mechanical torque | 0.780Nm |
| K | Torque constant | 0.433Nm/A |
| R | Armature resistance | 3.15Ohm |
| L | Armature inductance | 8.15mH |
| $f_\nu$ | The friction forces | 0.0172Nms/rad |

The experimental test is considered in a period of time which not exceeds 100 s.

Figures 6 and 7 show that, using the second order sliding mode control we can reach the steady state in a short time (10s). Also, for the first test, we have $\theta_1$=0.58 deg, $\theta_2$=0.68 deg, from that we notice that the azimuth inclination $\theta_1$ is very close to the altitude inclination $\theta_2$. This fact is related to the elliptic movement nature of the sun. The real and observed angular velocities (Fig.8 and Fig.9) of the two axis tracker are very close: $\Omega_1$=180 deg/s $\approx \hat{\Omega}_1$ and $\Omega_2$=160 deg/s $\approx \hat{\Omega}_2$. This excellent result is due to the high estimation quality of the sliding mode observer. Moreover, the SMC shows its robustness in Fig.5 and 6 when we inject a periodic disturbance signal (Fig.12) which the signal period *T=15s*. We see that the system has lightly deviate from trajectory then come back to the equilibrium position. In another hand, the control evolution (Fig.10), with no high level and no sharp commutation frequency, can give good operating conditions for the actuators (both azimuth and altitude motors).

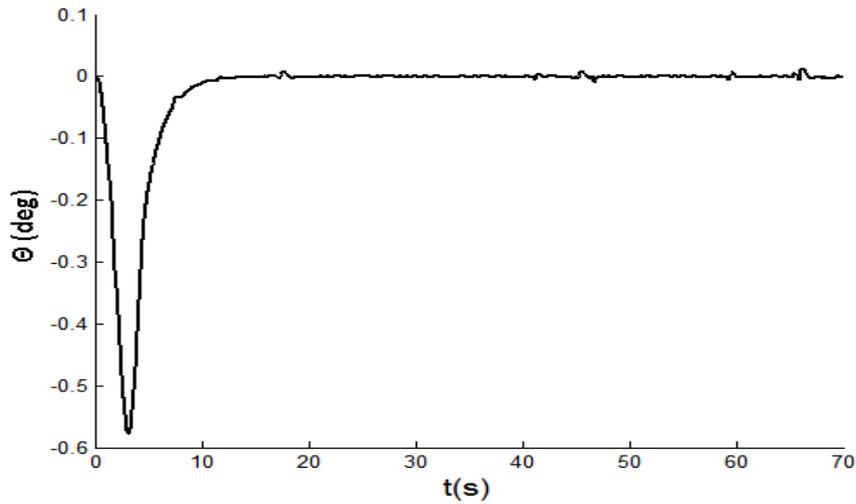

Figure 6.  Inclination angle of the azimuth motor

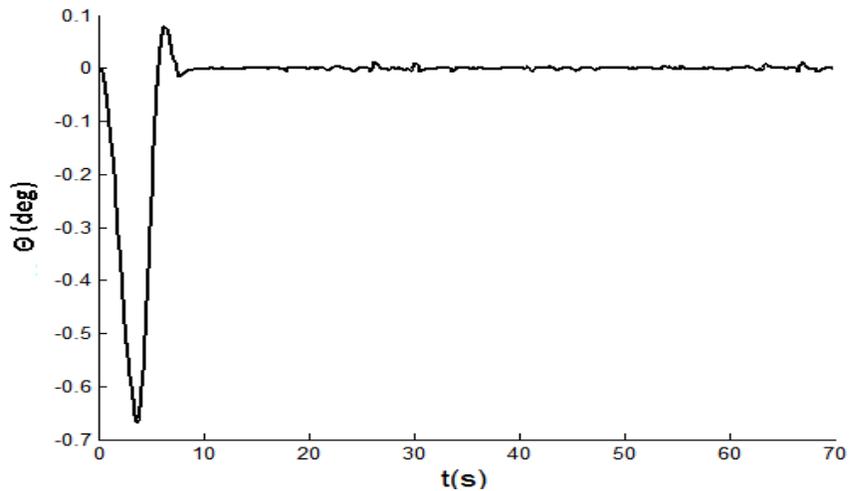

Figure 7. Inclination angle of the altitude motor

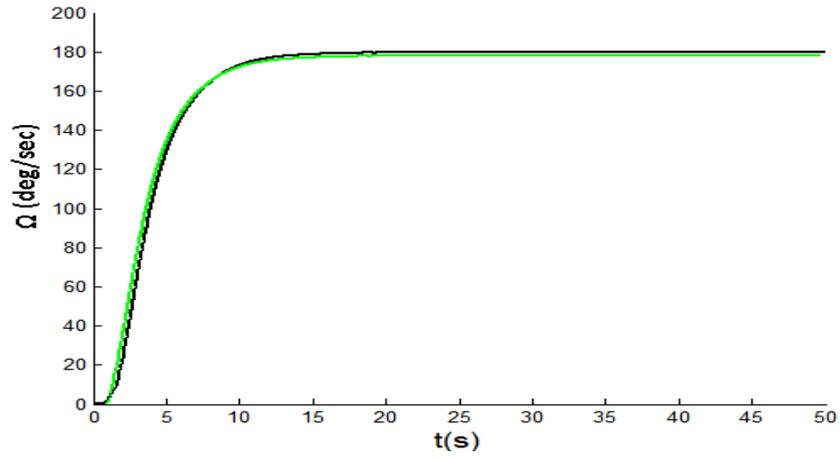

Figure 8. Real and observed Angular velocity of the azimuth motor

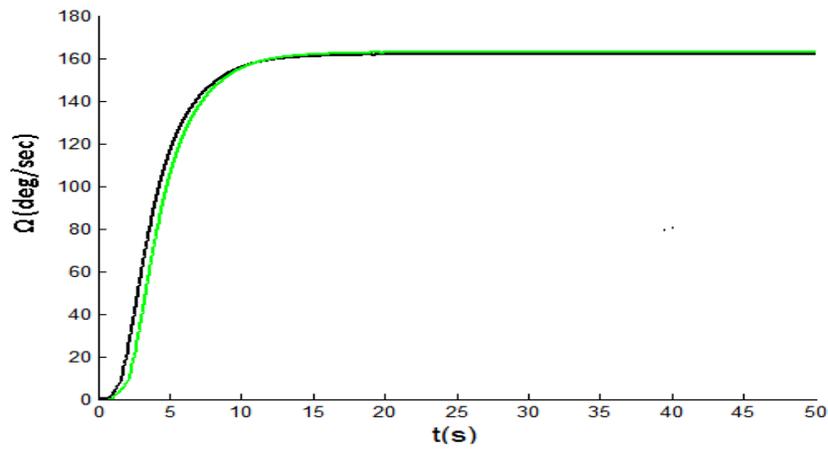

Figure 9. Real and observed Angular velocity of the altitude motor

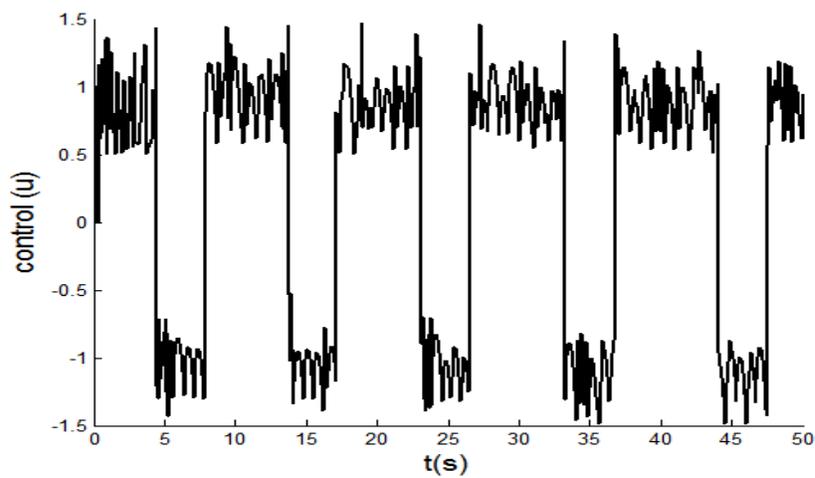

Figure 10. System control evolutions

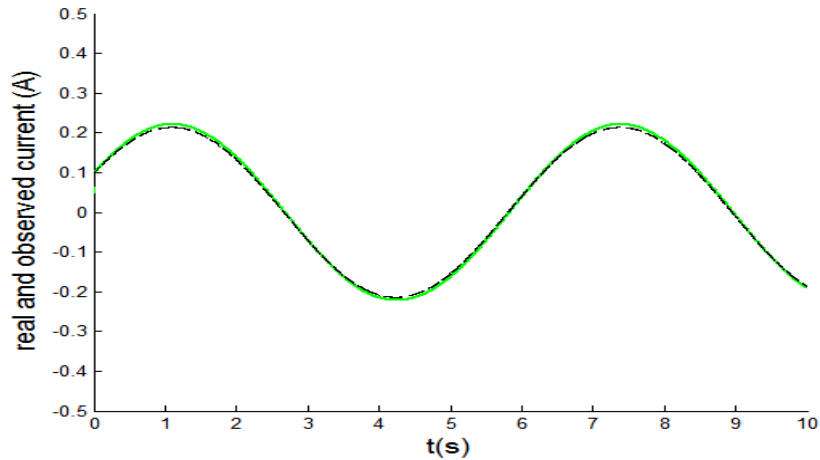

Figure 11. The real and observed current of the control

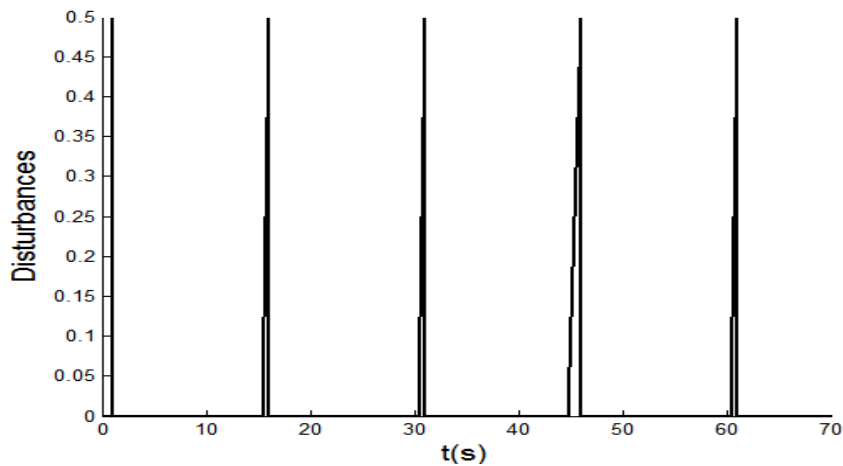

Figure 12. Disturbance signal

## 6. CONCLUSIONS

In this work, we approached a review of the literature of tracking process for the dual axis sun tracker by a sliding mode control law. In this way, different controllers have been presented. The synthesis of a control law by second order sliding mode using a nonlinear sliding surface has been treated. Then a sliding mode observer has been considerate to evaluate the angular velocities of the stepper motors. Experimental results show the effectiveness of the sliding mode control in the tracking process and its robustness unless disturbances and the high estimation quality of the sliding mode observer.

## Author

**Ahmed Rhif** was born in Sousah, Tunisia, in August 1983. He received his Engineering diploma and Master degree, respectively, in Electrical Engineering in 2007 and in Automatic and Signal Processing in 2009 from the National School of Engineer of Tunis, Tunisia (L'Ecole Nationale d'Ingénieurs de Tunis E.N.I.T). He has worked as a Technical Responsible and as a Project Manager in both LEONI and CABLITEC (Engineering automobile companies). Then he has worked as a research assistant at the Private University of Sousah (Université Privée de Sousse U.P.S) and now in the High Institute of Applied Sciences and Technologies of Sousah (Institut Supérieur des Sciences Appliquées et de Technologie de Sousse I.S.S.A.T.so). He is currently pursuing his PhD degree in the Polytechnic School of Tunis (E.P.T). His research interest includes control and nonlinear systems. **E-mail:** ahmed.rhif@gmail.com

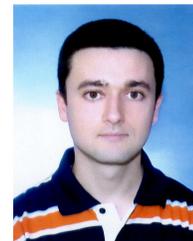